\documentclass[%
 reprint,
 amsmath,amssymb,
 aps,
pra,
]{revtex4-2}

\usepackage{graphicx}
\usepackage{dcolumn}
\usepackage{bm}
 \usepackage{latexsym}
 \usepackage{amsmath}
 \usepackage{amsfonts}
 \usepackage{amssymb}
 \usepackage{subfigure}
 \usepackage{color}
 \usepackage{setspace}
 \usepackage{titlesec}
 \usepackage{mathtools}

 \newcommand{\bc}{\begin{center}}
 \newcommand{\ec}{\end{center}}

\begin{document}

\preprint{APS/123-QED}

\title{Coherent control of levitated nanoparticles via dipole-dipole interaction}

\author{Sandeep Sharma}
 \email{quanta.sandeep@gmail.com}
\affiliation{Department of Physics, KAIST, Daejeon 34141, Republic of Korea}%
\author{Seongi Hong}
\affiliation{Department of Physics, KAIST, Daejeon 34141, Republic of Korea}
\author{Andrey S. Moskalenko}
 \email{moskalenko@kaist.ac.kr}
\affiliation{Department of Physics, KAIST, Daejeon 34141, Republic of Korea}%

\date{\today}

\begin{abstract}
We propose a scheme to create and transfer thermal squeezed states and random-phase coherent states in a system of two interacting levitated nanoparticles. In this coupled levitated system, we create a thermal squeezed state of motion in one of the nanoparticles by parametrically driving it and then transferring the state to the other nanoparticle with high fidelity. The transfer mechanism is based on inducing a non-reciprocal type of coupling in the system by suitably modulating the phases of the trapping lasers and the inter-particle distance between the levitated nanoparticles. This non-reciprocal coupling creates a unidirectional channel where information flows from one nanoparticle to the other nanoparticle but not vice versa, thereby allowing for transfer of mechanical states between the nanoparticles with high fidelity. We also affirm this transfer mechanism by creating and efficiently transferring a random-phase coherent state in the coupled levitated system. Further, we make use of the feedback nonlinearity and parametric driving to create simultaneous bistability in the coupled levitated system. Our results may have potential applications in quantum information processing, quantum metrology, and in exploring many-body physics under a controlled environment.
\end{abstract}

\maketitle


\section{\label{sec:level1}Introduction}

Optically levitated single nanoparticles, owing to their high tunability and low decoherence, have emerged as an ideal candidate for applications in high precession sensing \cite{Gieseler1}, testing fundamental limits in physics \cite{Romero-Isart1} and exploring non-equilibrium physics \cite{Gonzalez-Ballestero1}. Further, with recent developments towards trapping of multiple nanoparticles, a new avenue has been opened up which holds potential for exploration of coupled dynamics using these levitated nanoparticle arrays \cite{Yan1, Arita1, Rieser1, Slezak1}. To this end, efforts have been made to study the optical binding interaction and Coulomb interaction between two levitated nanoparticles \cite{Arita1, Rieser1, Slezak1}, simultaneous cooling of the mechanical motion of two levitated nanoparticles \cite{Vijayan1, Penny1, Bykov1, Liska1}, manipulating coupling interaction between the nanoparticles\cite{Vijayan2}, entanglement dynamics \cite{Rudolph1, Chauhan1}, quantum correlations between levitated nanoparticles \cite{Brandao1}, differential force sensing \cite{Rudolph2}, and Hermitian and non-Hermitian physics \cite{Yokomizo1, Reisenbauer1, Liska2}.

In this work, we mainly focus on the coherent dynamics of optically interacting two levitated nanoparticles. Optically interacting levitated systems (OILSs) are highly tunable and hold a greater advantage over systems interacting via the Coulomb force as concerns their utilization as a platform for exploring various many-body physics. This is because the coupling strength between the levitated nanoparticles in OILS can be efficiently controlled by modulating the intensity of trapping lasers, the phases of trapping lasers, and the inter-particle distance\cite{Rieser1}, while such a high degree of control is limited in the systems interacting via coulomb force \cite{Slezak1}. Manipulation of coupling strength is an important aspect in the exploration of different quantum phenomena arising in the coupled systems \cite{Groblacher1, Verhagen1}. Towards such manipulation of coupling in OILS, an interesting feature known as a unidirectional coupling \cite{Xu1} arises by suitably modulating the trapping laser phases and the inter-particle distance. This unidirectional coupling opens a one-way channel of energy flow and can be used to transfer different mechanical states in these OILSs, which is a necessary factor in quantum information processing. However, to date, a detailed discussion of this transfer mechanism in OILs has remained unexplored. Achieving efficient mechanical state transfer in OILs can be an important development towards utilizing these systems for practical quantum information processing.

Motivated by this, we study the creation of different mechanical states in two optically interacting levitated nanoparticles, as considered in Ref.\cite{Rieser1}, and explore the possibility of transferring such states from one of the nanoparticles to another one. We start by creating a thermal squeezed state of motion in one of the nanoparticles via a parametric drive. Next, by inducing a unidirectional coupling in the system, we find that the thermal squeezed state can be transferred to the other nanoparticle with very high fidelity. Further, to demonstrate the effectiveness of our transfer scheme, we also study the creation and transfer of a random-phase coherent state in the coupled system and find a high-fidelity induced transfer rate for this case as well. Finally, we utilize this unidirectionality phenomenon to generate simultaneous bistability in the studied coupled levitated system.

\section{Theoretical Model}
\label{Theo.1}
\begin{figure}[ht!]
\centering\includegraphics[height=6.5cm, width=6.5cm]{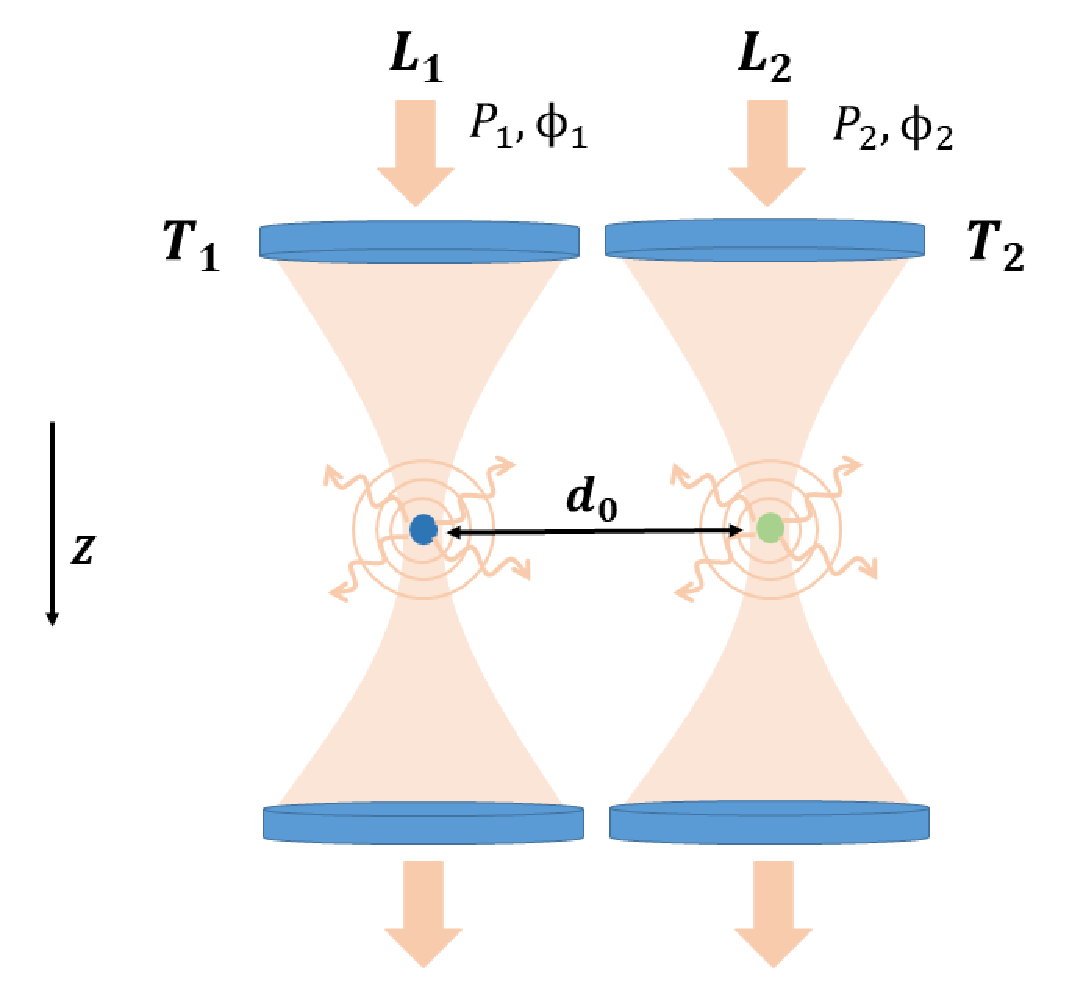}%
 \caption[]{A schematic diagram of a system of two interacting levitated nanoparticles. The nanoparticles labeled as blue and green dots are trapped at an inter-particle distance $d_{0}$ by two optical tweezers $T_{1}$ and $T_{2}$. Both nanoparticles interact with each other via photons scattered from trapping lasers $L_{1}$ and $L_{2}$ having powers $P_{1}$ and $P_{2}$ and phases $\phi_{1}$ and $\phi_{2}$, respectively.}
\label{fig:fig1}%
\end{figure}

We consider two dielectric nanoparticles having mass $m$ and radius $r$ that are levitated in different optical trap potentials created by two distinct optical tweezers, as shown in Fig.~\ref{fig:fig1}. Both nanoparticles interact with each other via the scattered fields from one another, giving rise to a non-reciprocal type of optical binding force between them \cite{Rieser1}. To a good approximation, the trapping potentials can be considered harmonic \cite{Gieseler1}, and hence the two levitated nanoparticles can be seen as two interacting harmonic oscillators (HOs). 

The quantum dynamics of these interacting HOs can be captured by the following master equation \cite{Sharma1, Rudolph3}:
\begin{align}
\label{Eq_1}
\dot{\rho}&=\sum_{j=1}^{2}-i[(\omega_{j}b^{\dagger}_{j}b_{j}+ \frac{S_{j}}{4}Q^{2}_{z_{j}}),\rho]-\sum_{j=1}^{2} i\frac{\gamma_{gj}}{2}[Q_{z_{j}},\{P_{z_{j}},\rho\}] \nonumber \\
                           &-\sum_{j=1}^{2} \frac{D_{tj}}{2}\mathcal{D}[Q_{z_{j}}]\rho+ \sum_{\substack{j,j'=1 \\ j\neq j'}}^{2}\frac{iS_{jj'}}{2}[Q_{z_{j}},\{Q_{z_{j'}},\rho\}],
\end{align}
where $\rho$ represents the two-particle density matrix for the coupled levitated system. The dimensionless position and momentum operators for the nanoparticles are designated by $Q_{z_{j}}=(b^{\dagger}_{j}+b_{j})$ and $P_{z_{j}}=i(b^{\dagger}_{j}-b_{j})$, respectively, with $b^{\dagger}_{j}$ and $b_{j}$ denoting the phonon creation and annihilation operators of the mechanical mode of the nanoparticles. Further, both $b^{\dagger}_{j}$ and $b_{j}$ obey the bosonic commutation relation $[b_{j},b^{\dagger}_{j'}]=\delta_{jj'}$. The Lindblad superoperator $\mathcal{D}[\mathcal{O}]$ acts on $\rho$ as follows: $\mathcal{D}[\mathcal{O}]\rho=\mathcal{O}^{\dagger}\mathcal{O}\rho+\rho\mathcal{O}^{\dagger}\mathcal{O}-2\mathcal{O}^{\dagger}\rho\mathcal{O}$.

The first term on the right-hand side (rhs) of Eq.~(\ref{Eq_1}) consists of two parts. The first part corresponds to the harmonic motion of both nanoparticles with frequencies $\omega_{1}$ and $\omega_{2}$, respectively. The second part represents the effect of non-reciprocal binding force on the motion of the nanoparticle with strengths $S_{1}$ and $S_{2}$, respectively. The second term on the rhs of Eq.~(\ref{Eq_1}) reflects the damping of nanoparticle motion with a rate $\gamma_{gj}$ due to the surrounding gas. The third term represents the effect of photon scattering ($A_{tj}$) and gas scattering ($D_{pj}$) on the motion of nanoparticle with the total rate $D_{tj}= A_{tj}+ D_{pj}$ \cite{Pettit1}. The fourth term is responsible for non-reciprocal coupling between the nanoparticles with rates $S_{jj'}$. The strengths of the optical binding force, $S_{1}$ and $S_{2}$, depend on the inter-particle distance $d_{0}$ and relative phase difference between trapping lasers $\Delta \phi=\phi_{1}-\phi_{2}$ and are given by $S_{1}=g_{1}+g_{2}$ and $S_{2}=g_{1}-g_{2}$, with $g_{1}=g$ cos($kd_{0}$)cos($\Delta \phi)/kd_{0}$ and $g_{2}=g$ sin($kd_{0}$)sin($\Delta \phi)/kd_{0}$. Here $g$ is the modulating constant given by $g=\alpha^{2}k^{3}(k-1/z_{R})^{2}\sqrt{P_{1}P_{2}}/(2c\omega^{2}_{b}\pi^{2}\epsilon^{2}_{0})$, where $\alpha$ is the polarizability of nanoparticles, $k$ is the wave vector of trapping lasers, $P_{1}(P_{2})$ is the power of the trapping laser beam 1 (2), $\omega_{b}$ is the beam waist, $c$ is the speed of light, and $\epsilon_{0}$ is the vacuum permittivity \cite{Rieser1}. Further, the coupling strengths can be expressed as $S_{12}=g$ cos($kd_{0}-\Delta \phi$)$/kd_{0}$ and $S_{21}=g$ cos($kd_{0}+\Delta \phi$)$/kd_{0}$.

At first, we try to gain a basic understanding of the dynamics of a system of two interacting levitated nanoparticles in the absence of any kind of external drive or feedback. For this, we analyze the stochastic equations of motion (EOMs) for the coupled levitated system, which are described as
\begin{align}
\label{Eq_2a}
\ddot{Q}_{z_{1}}&=-\omega^{2}_{1}Q_{z_{1}}-2\gamma_{g1}\dot{Q}_{z_{1}}-\omega_{1}S_{1}Q_{z_{1}} \nonumber \\
&+\omega_{1}S_{12}Q_{z_{2}}+\omega_{1}\mathcal{F}_{1}, \\
\ddot{Q}_{z_{2}}&=-\omega^{2}_{2}Q_{z_{2}}-2\gamma_{g2}\dot{Q}_{z_{2}}-\omega_{2}S_{2}Q_{z_{2}}\nonumber \\
&+\omega_{2}S_{21}Q_{z_{1}}+\omega_{2}\mathcal{F}_{2},
\label{Eq_2b}
\end{align}
where $\mathcal{F}_{1}$ and $\mathcal{F}_{2}$ are the Langevin forces acting on nanoparticle 1 and 2, respectively. A detailed derivation of the above EOMs is provided in the Appendix [see~\ref{AppxB}].
It is evident from the expressions for $S_{21}$ and $S_{12}$ that for $kd_{0}=2n\pi+\pi/4$ and $\Delta \phi=2n\pi+\pi/4$, with $n$ being a non-negative integer, Eq.~(\ref{Eq_2b}) becomes independent of $Q_{z_{1}}$, while Eq.~(\ref{Eq_2a}) remains dependent on $Q_{z_{2}}$. This implies that for the aforementioned condition, a unidirectional coupling is induced in the system, where there is an energy flow from particle 2 to particle 1 but not vice versa \cite{Xu1}. Note that in this case we also have $S_{2}=0$. In the next section, we will utilize this unidirectionality phenomenon and explore the possibility of transferring different mechanical states from one particle to another. 


\section{State creation and transfer}
In this section, we demonstrate the creation of a squeezed thermal state and a random-phase coherent state in the mechanical mode of nanoparticle 2 as well as their transfer to nanoparticle 1 via a unidirectional coupling with high fidelity. Additionally, we also show simultaneous bistable dynamics in both levitated nanoparticles due to this coupling.

\subsection{Squeezed state}
\label{squez}
\begin{figure}[t!]
\centering\includegraphics[height=6.5cm, width=7.5cm]{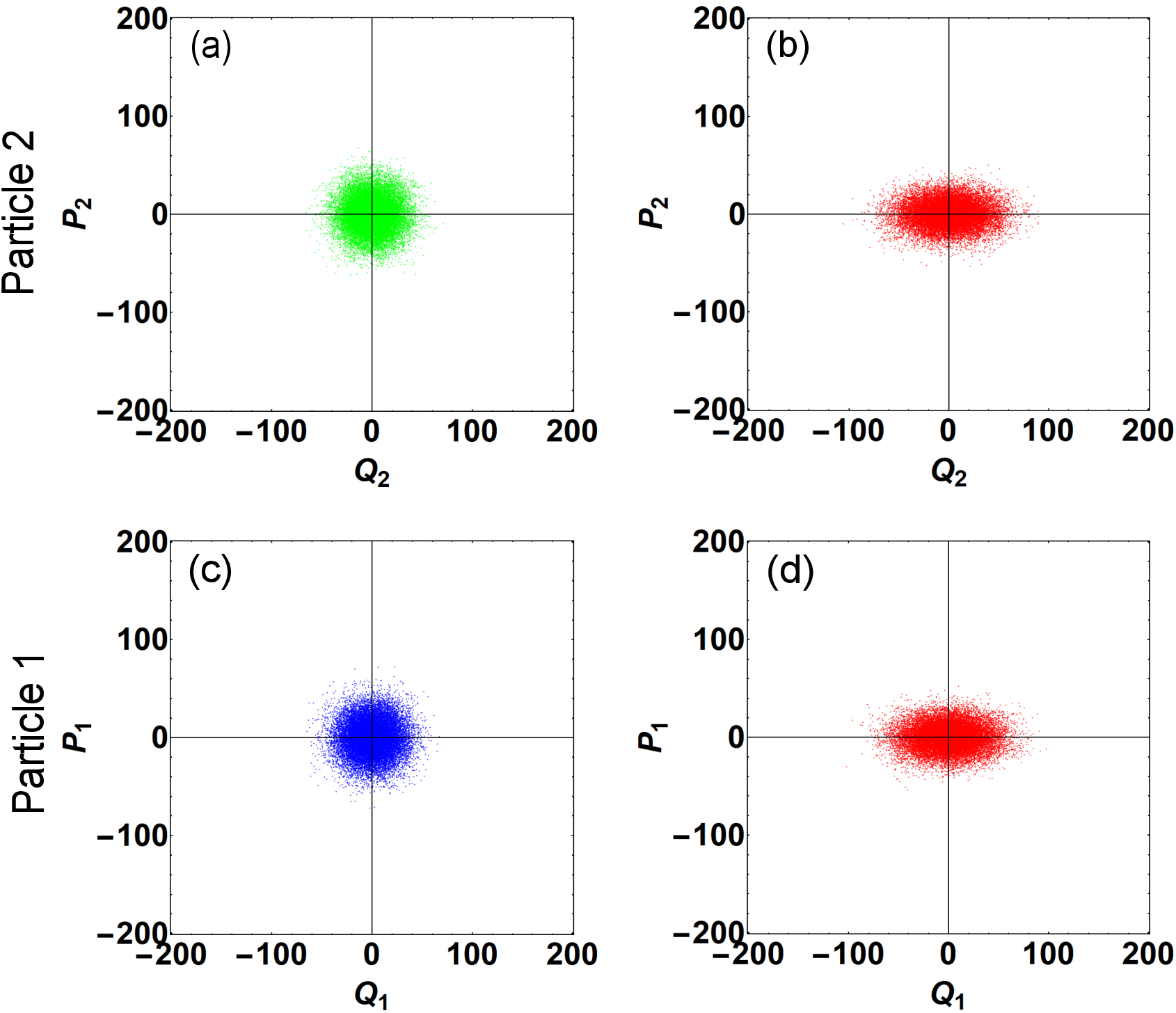}%
 \caption[]{Projected motion of both nanoparticles in phase-space. Panel (a) [(b)] shows the initial [final] state of motion of nanoparticle 2. Panel (c) [(d)] shows the state of nanoparticle 1 before [after] a unidirectional coupling. Parameters: $\omega_{0}=127~$KHz, $A_{t}=1~$KHz, $\gamma_{g}=1~$Hz, $\gamma_{a}=0~$Hz, $r=0.8$, $s=100~$Hz, and $\gamma_{f}=0~$Hz, corresponding to experimental values from Refs.\cite{Rieser1, Pettit1}.}
\label{fig:fig2}%
\end{figure}

To create a squeezed state of motion for nanoparticle 2, we parametrically drive it with a force having strength $f\omega^{2}_{2}$ and tuned at twice the oscillation frequency of the nanoparticle \cite{Bothner1}.  Under the action of this force and with the condition $S_{2}=S_{21}=0$, the EOMs for the levitated nanoparticles can be written as
\begin{align}
\label{Eq_3a}
\ddot{Q}_{z_{1}}&=-\omega^{2}_{1}Q_{z_{1}}-\gamma_{g1}\dot{Q}_{z_{1}} \nonumber \\
&-2sQ_{z_{1}}+2sQ_{z_{2}}+\mathcal{F}_{1}, \\
\ddot{Q}_{z_{2}}&=-\omega^{2}_{2}Q_{z_{2}}-\gamma_{g2}\dot{Q}_{z_{2}} \nonumber \\
&-f\omega^{2}_{2}\sin(2\omega_{2}t)Q_{z_{1}}+\mathcal{F}_{2},
\label{Eq_3b}
\end{align}
where $s=g/kd_{0}$ [see~\ref{AppxB}]. For simplicity, we then assume that both nanoparticles have the same frequency $\omega_{1}=\omega_{2}=\omega_{0}$ and are subjected to the same damping $\gamma_{g1}=\gamma_{g2}=\gamma_{g}$. We also assume equal scattering rates $A_{t1}=A_{t2}=A_{t}$ for both nanoparticles. This situation can be easily created in experiments by controlling the trapping laser intensity and the gas pressure \cite{Rieser1, Pettit1}. To visualize the creation and transfer of thermal squeezed states, we study the phase-space dynamics of the coupled levitated system by making the following ansatz for the solution of Eq.~(\ref{Eq_3a}) and (\ref{Eq_3b}):
\begin{align}
\label{Eq_4}
Q_{z_{j}}=Q_{j}~\cos(\omega_{0}t)+P_{j}~\sin(\omega_{0}t), ~~~~j = 1,2.
\end{align}
Here $Q_{j}$ and $P_{j}$ are slowly varying quadrature components of the motion of the nanoparticles \cite{Mahboob1, Pontin1}.
Next, by utilizing Eq.~(\ref{Eq_4}) in Eqs.~(\ref{Eq_3a}) and (\ref{Eq_3b}), we can write the EOMs for the quadrature components of both nanoparticles as
\begin{align}
\label{Eq_6a}
\dot{Q}_{1}&=-\gamma_{g}Q_{1}+sP_{1}-sP_{2}-\frac{\mathcal{F}_{s1}}{2}, \\
\dot{P}_{1}&=-\gamma_{g}P_{1}-sQ_{1}+sQ_{2}+\frac{\mathcal{F}_{c1}}{2}, \\
\dot{Q}_{2}&=-\left(\gamma_{g}-r\gamma_{g}\right)Q_{2}-\frac{\mathcal{F}_{s2}}{2}, \label{Eq_6b} \\
\dot{P}_{2}&=-\left(\gamma_{g}+r\gamma_{g}\right)P_{2}+\frac{\mathcal{F}_{c2}}{2}, 
\label{Eq_6c}
\end{align}
where $r=f\omega_{0}/\gamma_{g}$ is the squeezing strength and $\mathcal{F}_{jc}$ and $\mathcal{F}_{js}$ are slowly varying cosine and sine components of the Langevin forces, respectively [for details, see~\ref{AppxB}]. To numerically solve Eqs.~(\ref{Eq_6a})-(\ref{Eq_6c}), we follow the approach as in Ref.\cite{Mahboob1} and present the solution obtained for the long-time interaction limit in the form of phase-space plots as shown in Fig.~(\ref{fig:fig2}). In Fig.~(\ref{fig:fig2}), panels (a) and (b) show the phase-space distribution of the motion of nanoparticle 2 before and after the parametric driving, respectively. Fig.~\ref{fig:fig2}(a) shows a circularly symmetric distribution indicating that nanoparticle 2 is in a thermal state, which is expected as initially the nanoparticle motion is solely driven by thermal Langevin forces \cite{Mahboob1}. From Fig.~\ref{fig:fig2}(b), it is evident that when a parametric drive is applied, the thermal fluctuations along one of the quadrature components are amplified, while along the other component, they are de-amplified, resulting in a squeezed distribution. This reflects the creation of a thermal squeezed state of motion of nanoparticle 2. To quantify the squeezing induced in the system, we evaluate the variances of the quadrature components of the motion of nanoparticle 2, given by
\begin{align}
\label{Eq_7a}
\sigma^{2}_{Q_{2}}&=\frac{1}{\pi}\int_{-\infty}^{\infty} \langle |Q_{2}(\omega)|^{2}\rangle d\omega = \frac{A_{t}}{4\gamma_{g}(1-r)}, \\
\sigma^{2}_{P_{2}}&=\frac{1}{\pi}\int_{-\infty}^{\infty} \langle |P_{2}(\omega)|^{2}\rangle d\omega = \frac{A_{t}}{4\gamma_{g}(1+r)}. 
\label{Eq_7b}
\end{align}
Here $Q_{2}(\omega)$ and $P_{2}(\omega)$ are derived by solving Eqs.~(\ref{Eq_6b}) and (\ref{Eq_6c}) in the frequency domain \cite{Mahboob1, Pontin1, Aspelmeyer1}. From Eqs.~(\ref{Eq_7a}) and (\ref{Eq_7b}) it is apparent that when the squeezing strength $r$ vanishes, the variances of both quadrature components become equal. This indicates that the phase-space distribution has to be symmetric, which is in accordance with the result shown in Fig.~\ref{fig:fig2}(a). For $r\ne 0$, the variance of quadrature $Q_{2}$ is greater than that of the $P_{2}$, corresponding to a squeezed distribution in the phase space, as shown in Fig.~\ref{fig:fig2}(b).
Next, we initiate a unidirectional coupling and study the phase-space dynamics of the motion of nanoparticle 1. It is clear from Fig.~\ref{fig:fig2} that in the presence of a unidirectional coupling, the state of nanoparticle 1 evolves from a thermal state, as shown in Fig.~\ref{fig:fig2}(c), to a squeezed state revealed in Fig.~\ref{fig:fig2}(d). Further, following a similar approach as mentioned above, we quantify squeezing in the motion of nanoparticle 1 by deriving the variances of its quadrature components. Assuming that both nanoparticles interact for a long time, eventually reaching a steady state, under the condition $s>\gamma_{g}$ we write the variances as
\begin{align}
\label{Eq_8a}
\sigma^{2}_{Q_{1}}&=\frac{1}{\pi}\int_{-\infty}^{\infty} \langle |Q_{1}(\omega)|^{2}\rangle d\omega \nonumber \\
&\approx \frac{A_{t}}{4\gamma_{g}(1-r^{2})s^{2}}\left[\gamma^{2}_{g}(1-r)+s^{2}(1+r)\right] \nonumber \\
&\approx \frac{A_{t}}{4\gamma_{g}(1-r)}~~~~ \mbox{for} ~~ s \gg \gamma_{g}, \\
\sigma^{2}_{P_{1}}&=\frac{1}{\pi}\int_{-\infty}^{\infty} \langle |P_{1}(\omega)|^{2}\rangle d\omega  \nonumber \\
&\approx \frac{A_{t}}{4\gamma_{g}(1-r^{2})s^{2}}\left[\gamma^{2}_{g}(1+r)+s^{2}(1-r)\right] \nonumber \\
&\approx \frac{A_{t}}{4\gamma_{g}(1+r)}~~~~ \mbox{for} ~~ s \gg \gamma_{g}.
\label{Eq_8b}
\end{align}
From the analysis of both the phase space distribution and the variances of the quadrature components of both nanoparticles, we can confirm that a unidirectional coupling induces a transfer of mechanical states from nanoparticle 2 to nanoparticle 1. Additionally, to verify the validity of our numerical results on the induced state transfer, we compare them with corresponding analytical results obtained by solving the Fokker-Planck equation in the steady-state limit [see~\ref{AppxC} \& \ref{AppxD}]. Towards this, we found excellent agreement between the numerical and analytical results on the phase-space distribution [see Fig.~\ref{fig:fig6}] and the variances [see Eqs.~(\ref{Eqn_A16})-(\ref{Eqn_A17})], which affirms the validity of our analysis elucidating such state transfer mechanism in coupled levitated systems. Further, to determine the efficiency of this induced mechanical state transfer process, we find the transfer fidelity by using Eq.~(\ref{Eqn_A18}). In our case, by considering achievable experimental values of parameters \cite{Rieser1, Pettit1}, we found the fidelity $\mathcal{F}=0.999$, which is better than the value achieved in other optomechanical systems \cite{Neto1, Xi1, Navarathna1}. 
\subsection{Random-phase Coherent state}
\begin{figure}[t!]
\centering\includegraphics[height=4.5cm, width=9.0cm]{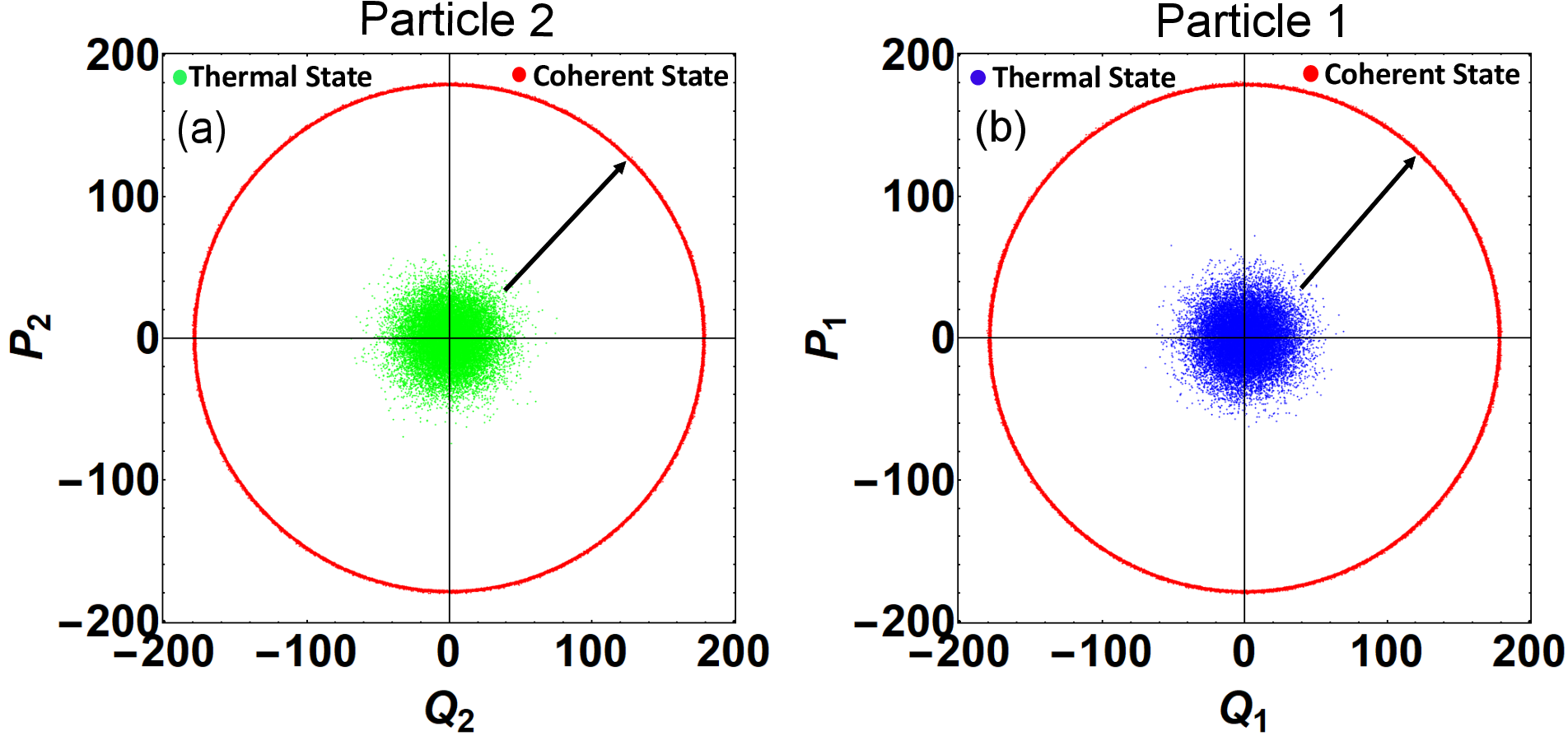}%
 \caption[]{Induced transfer of a random-phase coherent state between the nanoparticles of the coupled levitated system. Panel (a) [(b)] depicts the initial thermal state and final random-phase coherent state of motion of nanoparticle 2 [1]. The arrows indicate the time evolution of the states of nanoparticles. Parameters: $\gamma_{a}=20~$Hz, $r=0$, and $\gamma_{f}=10^{-4}~$Hz. Other parameters are the same as in Fig.~\ref{fig:fig2}.}
\label{fig:fig3}%
\end{figure}

\begin{figure}[t!]
\centering\includegraphics[width=7.5cm]{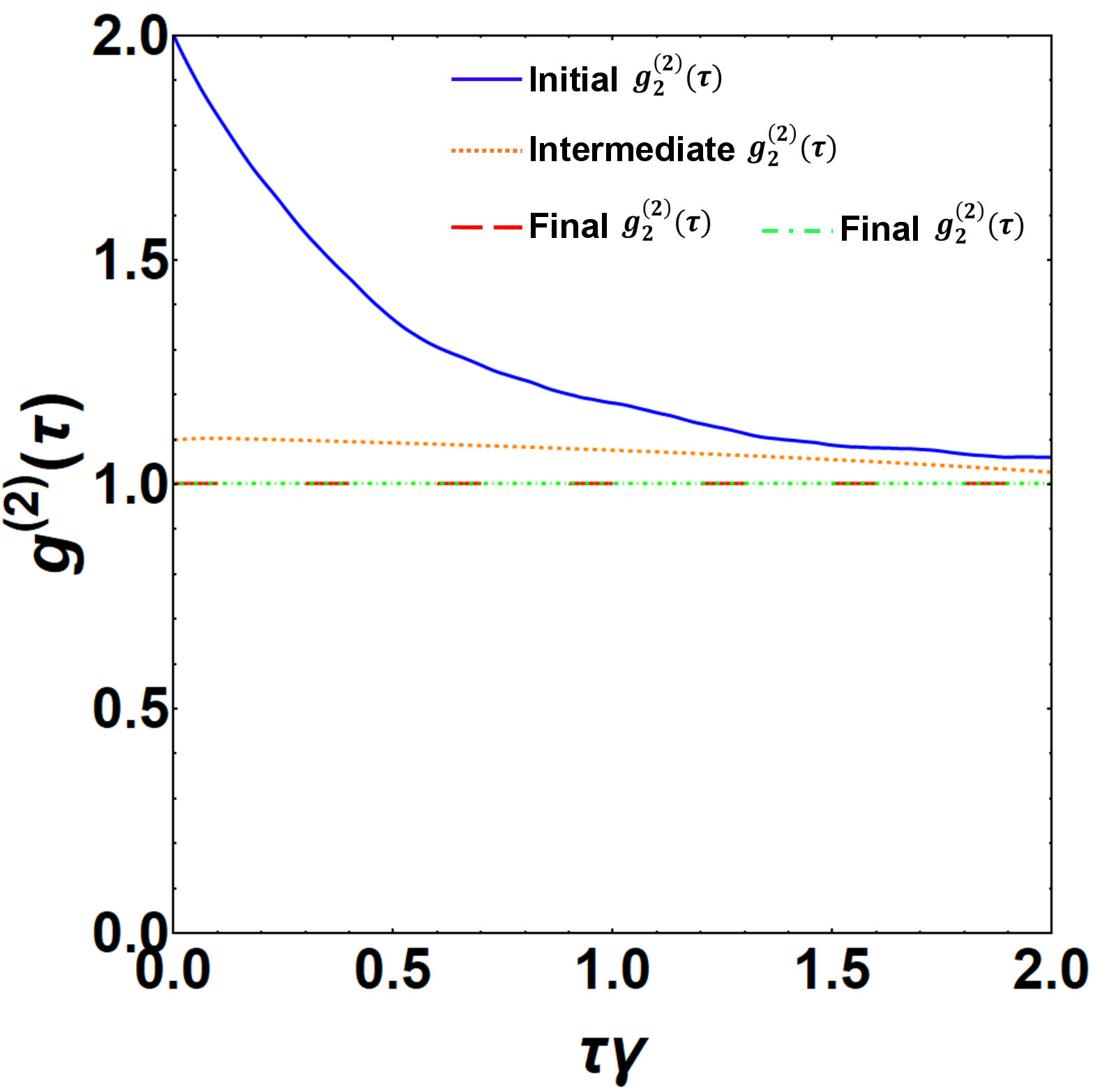}%
 \caption[]{Second-order coherence $g^{(2)}_{2}(\tau)$ $[g^{(2)}_{1}(\tau)]$ for nanoparticle 2 [1]. The blue solid [red long-dashed] line depicts the numerically evaluated $g^{(2)}_{2}(\tau)$ for the initial [final] state of nanoparticle 2.  $g^{(2)}_{2}(\tau)$ evaluated at an intermediate time moment $\tau=2/\gamma$ with an averaging time window of $4/\gamma$ during the evolution of nanoparticle 2 is represented by the orange dotted line. The green dotted-dashed line shows $g^{(2)}_{1}(\tau)$ for the final state of nanoparticle 1. The parameters considered here are the same as in Fig.~\ref{fig:fig3}.}
\label{fig:fig4}%
\end{figure}

In this section, to demonstrate that the coupled levitated system can be used for transferring other mechanical states, we also study the creation and transfer of random-phase coherent states \cite{Matsuo1} in this system. Similar to the above subsection, we first create a random-phase coherent state of motion in nanoparticle 2 and then transfer it to nanoparticle 1 via a unidirectional coupling. To create a random-phase coherent state of motion in nanoparticle 2, we simultaneously use linear feedback heating and non-linear feedback cooling to control its motional dynamics. To understand the effect of both these feedbacks on the motion of nanoparticle 2, we numerically solve the following EOMs for the quadrature components of the nanoparticles [see~\ref{AppxB} for details]
\begin{align}
\label{Eq_9a}
\dot{Q}_{1}&=-\gamma_{g}Q_{1}+sP_{1}-sP_{2}-\frac{\mathcal{F}_{s1}}{2}, \\
\dot{P}_{1}&=-\gamma_{g}P_{1}-sQ_{1}+sQ_{2}+\frac{\mathcal{F}_{c1}}{2}, \label{Eq_9b} \\
\dot{Q}_{2}&=-\left(\gamma_{g}-\gamma_{a}+6\gamma_{f}[Q^{2}_{2}+P^{2}_{2}]\right)Q_{2}-\frac{\mathcal{F}_{s2}}{2}, \label{Eq_9c} \\
\dot{P}_{2}&=-\left(\gamma_{g}-\gamma_{a}+6\gamma_{f}[Q^{2}_{2}+P^{2}_{2}]\right)P_{2}+\frac{\mathcal{F}_{c2}}{2}. 
\label{Eq_9d}
\end{align}
The result of our numerical solution is portrayed in the phase space as shown in Fig.~\ref{fig:fig3}. It can be seen in Fig.~\ref{fig:fig3}(a) that, in the presence of linear feedback heating and non-linear feedback cooling, the state of motion of nanoparticle 2 evolves from the thermal state to a random-phase coherent state. This is because the competition between gain (due to feedback heating) and loss (due to non-linear feedback cooling) drives the motion of nanoparticle 2 towards a stable oscillation. 
Next, to characterize the stable oscillation motion of nanoparticle 2, we also study the phonon population and second-order coherence. By analyzing Eq.~(\ref{Eq_9c}) and (\ref{Eq_9d}) in the steady state limit, we found that the phonon population saturates to a value $\approx (\gamma_{a}-\gamma_{g})/6\gamma_{f}$. We also studied the second-order coherence $g^{(2)}_{2}(\tau)$ using the approach as in Ref.\cite{Sharma1}. The results are illustrated in Fig.~\ref{fig:fig4}. It is clear that $g^{(2)}_{2}(\tau)$ evolves from a Lorentzian profile [for a thermal state] to a constant profile depicting nanoparticle 2 in a random-phase coherent state \cite{Matsuo1}. Thus, the above analysis on stable oscillation dynamics, phonon saturation effect, and second-order coherence together evidence the creation of a random-phase coherent state of motion in nanoparticle 2 \cite{Matsuo1}. 
Now, we switch on a unidirectional coupling and study the dynamics of nanoparticle 1 using Eqs.~(\ref{Eq_9a}) and (\ref{Eq_9b}) and present our numerical results in Fig.~\ref{fig:fig3}. It is evident from Fig.~\ref{fig:fig3}(b) that the motion of nanoparticle 1 also shows stable oscillation due to a unidirectional coupling. Further, the
second-order coherence $g^{(2)}_{1}(\tau)$ for the nanoparticle 1 converges to a constant profile, representing a random-phase coherent state, similar to that of nanoparticle 2, as shown in Fig.~\ref{fig:fig4}. With this, we can affirm the tunability of the induced state transfer mechanism in the studied coupled levitated system, which is enforced by a unidirectional coupling. Then, we also verify our numerically obtained results by comparing them with analytical results gained by solving the Fokker-Planck equation [see~\ref{AppxE} for details]. To this end, we also find excellent agreement between the outcomes of both approaches for the phase-space dynamics [see Fig.~\ref{fig:fig7} in~\ref{AppxE}] as well as for the phonon population [see Eq.~(\ref{Eqn_A21}) \& (\ref{Eqn_A22}) in~\ref{AppxE}]. Additionally, we check the efficiency of the transfer process by finding the corresponding fidelity using Eq.~(\ref{Eqn_A18}). Considering achievable experimental values of parameters, we could reach high fidelity $\mathcal{F}=0.999$ which is similar to that demonstrated in Sec.~\ref{squez} for the squeezing case. 

\subsection{Bistability}
\begin{figure}[htbp!]
\centering\includegraphics[height=4.5cm, width=9.0cm]{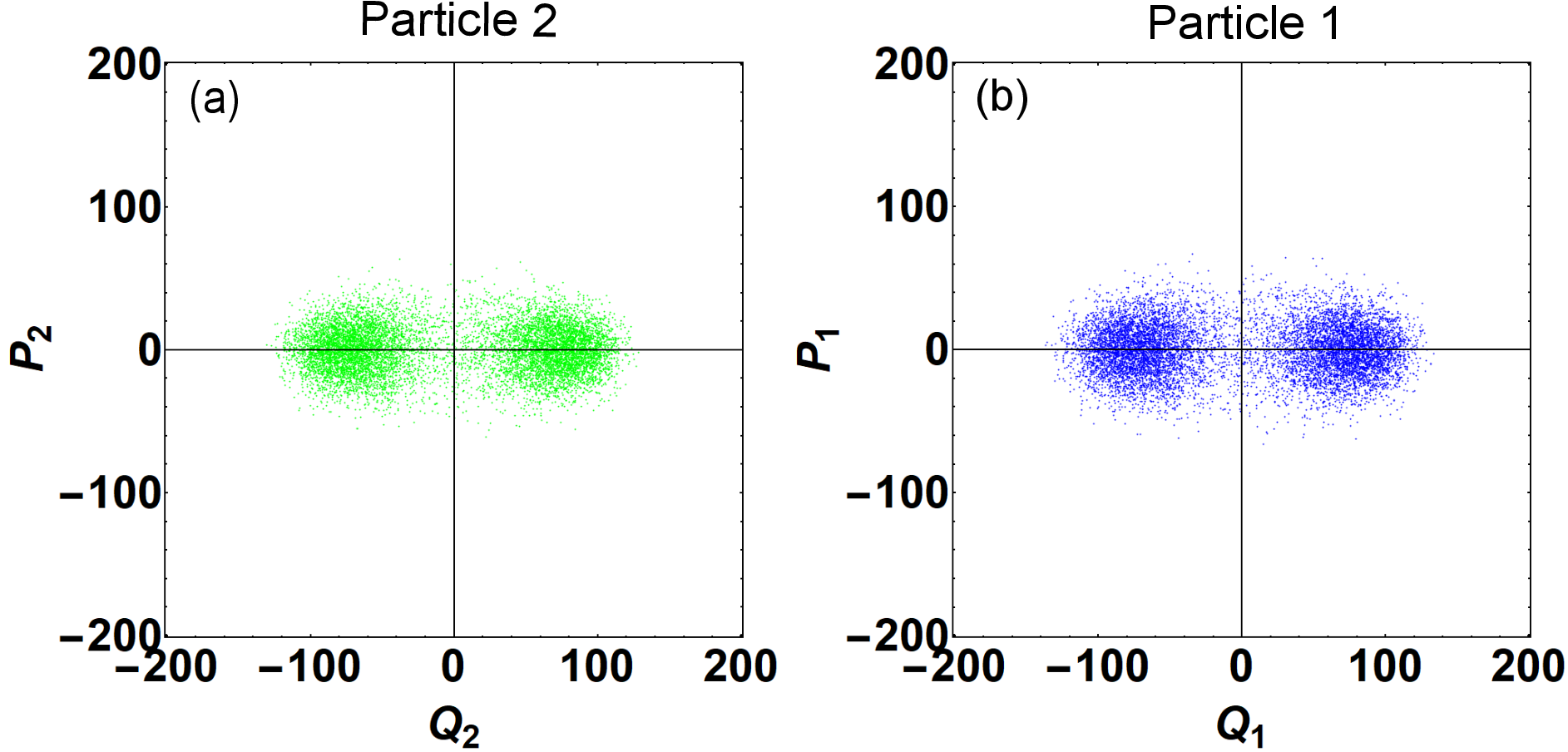}%
 \caption[]{Panel (a) [(b)] shows the bistable dynamics of nanoparticle 2 [1]. Parameters: $\gamma_{g1}=1$~Hz, $\gamma_{g2}=20$~Hz, $r=0.9$, and $\gamma_{f}=2\times10^{-4}$~Hz. Other parameters are the same as in Fig.~(\ref{fig:fig2}).}
\label{fig:fig5}%
\end{figure}

In this section, we make use of the unidirectionality phenomenon as a tool to initiate simultaneous bistable dynamics of the coupled levitated system. In this regard, we first apply non-linear feedback cooling to nanoparticle 2 and then use parametric driving to create bistable dynamics of the motion of the nanoparticle. 
Specifically, we numerically solve the EOMs for the quadrature components of the nanoparticles [see~\ref{AppxB} for details]:
\begin{align}
\label{Eq_10a}
\dot{Q}_{1}&=-\gamma_{g1}Q_{1}+sP_{1}-sP_{2}-\frac{\mathcal{F}_{s1}}{2}, \\
\dot{P}_{1}&=-\gamma_{g1}P_{1}-sQ_{1}+sQ_{2}+\frac{\mathcal{F}_{c1}}{2}, \label{Eq_10b} \\
\dot{Q}_{2}&=-\left(\gamma_{g2}-r\gamma_{g2}+6\gamma_{f}[Q^{2}_{2}+P^{2}_{2}]\right)Q_{2}-\frac{\mathcal{F}_{s2}}{2}, \label{Eq_10c} \\
\dot{P}_{2}&=-\left(\gamma_{g2}+r\gamma_{g2}+6\gamma_{f}[Q^{2}_{2}+P^{2}_{2}]\right)P_{2}+\frac{\mathcal{F}_{c2}}{2}. 
\label{Eq_10d}
\end{align}
The resulting phase-space distributions are presented in Fig.~\ref{fig:fig5}. It is apparent from Fig.~\ref{fig:fig5}(a) that in the presence of parametric drive and non-linear feedback cooling the motion of nanoparticle 2 shows two stable oscillations reflecting bistable dynamics. This bistability is caused by the formation of a double-well trapping potential, which can be controlled by manipulating the strength of the parametric drive as well as the non-linear feedback cooling rate \cite{Ricci2}. Next, by activating a unidirectional coupling channel, we observe that nanoparticle 1 also exhibits bistable dynamics, as shown in Fig.~\ref{fig:fig5}(b). 
\section{Conclusion}
In conclusion, we have explored the coherent dynamics in a system of two interacting levitated nanoparticles. We found that the interaction strength between the levitated nanoparticles can be efficiently controlled by changing the inter-particle distance and the phase difference between the trapping lasers. In particular, we showed that at specific values of the inter-particle distance and the phase difference, a unidirectional energy flow channel can be opened up in the coupled system. We performed numerical simulations on its phase-space dynamics and showed that a unidirectional coupling enables the transfer of different mechanical states between the nanoparticles. We also made analytical calculations to illuminate the underlying transfer mechanism and found it to be in excellent agreement with our results from numerical simulations. To estimate the efficiency of the induced state transfer process we determined the transfer fidelity and demonstrated that its value can be very high, even better than for clamped optomechanical systems. Finally, we also used a unidirectional coupling to induce simultaneous bistability in the studied system. We expect that our results on mechanical state transfer may have potential applications in quantum information processing \cite{Riedinger1} and sensing \cite{Degen1}. Additionally, our findings on long-lived simultaneous coherent oscillation can be extended to many-body systems wherein it may be interesting to study coherent quantum thermodynamics \cite{Kwon1, Manzano1} and quantum metrology \cite{Chu1}. Further, results on simultaneous bistable dynamics in coupled systems may motivate future explorations of many interesting nonequilibrium many-body dynamics such as quantum critical phenomena and phase transitions \cite{Landa1, Foss-Feig1}. 

\begin{acknowledgments}
This research was supported by the National Research Foundation of Korea (NRF) grants funded by the Korea Government (MSIT) (Grant No. 2022R1I1A1A01053604 and Grant No. 2020R1A2C1008500). A.S.M was supported by the Mercator Fellowship of the Deutsche Forschungsgemeinschaft (DFG) - Project No. 425217212 - SFB 1432. S.S. acknowledges the financial support from KAIST through the BK21 postdoctoral fellowship.
\end{acknowledgments}

\section*{Appendix}
\label{App}

\makeatletter
\renewcommand{\@seccntformat}[1]{%
  \csname the#1\endcsname
  \ifnum\pdfstrcmp{#1}{section}=0 :\fi
  \quad
}
\makeatother
  
 \renewcommand{\thesection}{Appendix \Alph{section}}
  \setcounter{equation}{0}

\renewcommand{\theequation}{A-\arabic{equation}}
  \setcounter{equation}{0}  
%
%

\section{Master Equation}
\label{AppxA}
Here we present the full master equation which describes the complete dynamics of the system of two interacting levitated nanoparticles. It reads
\begin{align}
\label{Eqn_A1}
\dot{\rho}&=\sum_{j=1}^{2}\left(-i[(\omega_{j}b^{\dagger}_{j}b_{j}+ \frac{S_{j}}{4}Q^{2}_{z_{j}}),\rho]-i\frac{\gamma_{gj}}{2}[Q_{z_{j}},\{P_{z_{j}},\rho\}]\right) \nonumber \\
                           &-\sum_{j=1}^{2} \frac{D_{tj}}{2}\mathcal{D}[Q_{z_{j}}]\rho+ \sum_{\substack{j,j'=1 \\ j\neq j'}}^{2}\frac{iS_{jj'}}{2}[Q_{z_{j}},\{Q_{z_{j'}},\rho\}]\nonumber \\
                           &-i\frac{f\omega_{2}}{4}~\sin(2\omega_{2}t)[Q^{2}_{z_{2}},\rho]+i\frac{\gamma_{a}}{2}[Q_{z_{2}},\{P_{z_{2}},\rho\}] \nonumber \\
                           &-i\gamma_{f}[Q^{3}_{z_{2}},\{P_{z_{2}},\rho\}].
\end{align}
The definition and effect up to the fourth term on the right-hand side (rhs) of the above equation have been explained in Sec.~\ref{Theo.1}. The fifth term on the rhs is the parametric drive term and is responsible for generating squeezing in the system. The sixth and seventh terms represent linear feedback heating of the system with rate $\gamma_{a}$ and nonlinear feedback cooling of the system with rate $\gamma_{f}$, respectively. They are responsible for creating a random-phase coherent state in the system. Further, in the absence of feedback heating, the parametric drive and non-linear feedback terms give rise to bistability in the system.
%
%
\renewcommand{\theequation}{B-\arabic{equation}}
  \setcounter{equation}{0}  
\section{Langevin equation}
\label{AppxB}
To study the dynamics of the system, we make use of  Eq.~(\ref{Eqn_A1}) and write the Langevin equations of motion as \cite{Sharma1}
\begin{align}
\label{Eqn_A2}
\dot{Q}_{z_{1}}&=\omega_{1}P_{z_{1}}, \\
\dot{P}_{z_{1}}&=-\omega_{1}Q_{z_{1}}-2\gamma_{g1}P_{z_{1}}-S_{1}Q_{z_{1}}+S_{12}Q_{z_{2}}+\mathcal{F}_{1}, \\
\dot{Q}_{z_{2}}&=\omega_{2}P_{z_{2}}, \\
\dot{P}_{z_{2}}&=-\omega_{2}Q_{z_{2}}-2(\gamma_{g2}-\gamma_{a}+6\gamma_{f}Q^{2}_{z_{2}})P_{z_{2}}\nonumber \\
                     &-f\omega_{2}~sin(2\omega_{2}t)Q_{z_{2}}-S_{2}Q_{z_{2}}+S_{21}Q_{z_{1}}+\mathcal{F}_{2},
\label{Eqn_A3}
\end{align}
where the Langevin forces $\mathcal{F}_{1}$ and  $\mathcal{F}_{2}$ act on nanoparticle 1 and nanoparticle 2, respectively, and are represented as $\mathcal{F}_{1}=\sqrt{2K_{B}T\gamma_{g1}/\hbar\omega_{1}}~\xi_{T1}+\sqrt{D_{t1}}~\xi_{S1}$ and  $\mathcal{F}_{2}=\sqrt{2K_{B}T\gamma_{g2}/\hbar\omega_{2}}~\xi_{T2}+\sqrt{D_{t2}}~\xi_{S2}$. Further, $\xi_{Tji}$ and $\xi_{Sj}$ are stochastic noises corresponding to the effects from environment and scattering, respectively. They possess the following correlation properties: $\langle \xi_{Tj}(t)\xi_{Tj}(t') \rangle = \delta(t-t')$ and $\langle \xi_{Sj}(t)\xi_{Sj}(t') \rangle = \delta(t-t')$.

We combine the four above ordinary differential equations of the first order into two equations of the second order, 
\begin{align}
\label{Eqn_A4}
\ddot{Q}_{z_{1}}&=-\omega^{2}_{1}Q_{z_{1}}-2\gamma_{g1}\dot{Q}_{z_{1}}-\omega_{1}S_{1}Q_{z_{1}} \nonumber \\
&+\omega_{1}S_{12}Q_{z_{2}}+\omega_{1}\mathcal{F}_{1}, \\
\ddot{Q}_{z_{2}}&=-\omega^{2}_{2}Q_{z_{2}}-2(\gamma_{g2}-\gamma_{a}+6\gamma_{f}Q^{2}_{z_{2}})\dot{Q}_{z_{2}}\nonumber \\
&-f\omega_{2}~sin(2\omega_{2}t)Q_{z_{2}} -\omega_{1}S_{2}Q_{z_{2}}\nonumber \\
&+\omega_{2}S_{21}Q_{z_{1}}+\omega_{2}\mathcal{F}_{2}.
\label{Eqn_A5}
\end{align}
As discussed in Sec.~\ref{Theo.1}, we consider $S_{2}=S_{21}=0$, which is a prerequisite condition for a unidirectional state transfer. By assuming that both nanoparticles have the same oscillation frequency $\omega_{0}$ and the same damping $\gamma_{g}$, we can write the solution of Eqs.~(\ref{Eqn_A4}) and (\ref{Eqn_A5}) as
\begin{align}
\label{Eqn_A6}
Q_{z_{j}}=Q_{j}~\cos(\omega_{0}t)+P_{j}~\sin(\omega_{0}t), ~~~~j = 1,2.
\end{align}
Here $Q_{j}$ and $P_{j}$ are slowly varying quadrature components of the motion of nanoparticles\cite{Mahboob1}.
Further, as the nanoparticles respond at around $\omega_{0}$, the Langevin forces can also be written into the following form 
\begin{align}
\label{Eqn_A7}
\mathcal{F}_{j}=\mathcal{F}_{jc}~\cos(\omega_{0}t)+\mathcal{F}_{js}~\sin(\omega_{0}t), ~~~~j =1,2,
\end{align}
where $\mathcal{F}_{jc}$ and $\mathcal{F}_{js}$ are slowly varying cosine and sine components, respectively \cite{Mahboob1}. 

Substituting Eqs.~(\ref{Eqn_A6}) and (\ref{Eqn_A7}) into Eqs.~(\ref{Eqn_A4}) and (\ref{Eqn_A5}) and neglecting the second-order derivatives and higher-order frequency terms, we write the equations for the quadrature components of the motion of nanoparticles as
\begin{align}
\label{Eqn_A8}
\dot{Q}_{1}&=-\gamma_{g}Q_{1}+sP_{1}-sP_{2}-\frac{\mathcal{F}_{s1}}{2}, \\
\dot{P}_{1}&=-\gamma_{g}P_{1}-sQ_{1}+sQ_{2}+\frac{\mathcal{F}_{c1}}{2}, \\
\dot{Q}_{2}&=-\left(\gamma_{g}-r\gamma_{g}-\gamma_{a}+6\gamma_{f}[Q^{2}_{2}+P^{2}_{2}]\right)Q_{2}-\frac{\mathcal{F}_{s2}}{2}, \\
\dot{P}_{2}&=-\left(\gamma_{g}+r\gamma_{g}-\gamma_{a}+6\gamma_{f}[Q^{2}_{2}+P^{2}_{2}]\right)P_{2}+\frac{\mathcal{F}_{c2}}{2}, 
\label{Eqn_A9}
\end{align}
where $r=f\omega_{0}/\gamma_{g}$ is the squeezing strength and $s=\omega_{0}g/kd_{0}$ is the coupling strength.
%
%
\renewcommand{\theequation}{C-\arabic{equation}}
  \setcounter{equation}{0}  
\section{Fokker-Planck Equation}
\label{AppxC}
 We make use of the unidirectionality condition $S_{2}=S_{21}=0$ and proceed to transform the master equation, as in Eq.~(\ref{Eqn_A1}), into a Fokker-Planck (FP) equation.
 
At first, we make a unitary transformation $W=e^{-i\sum_{j=1}^{2}\omega_{j}b^{\dagger}_{j}b_{j}t}$ and transfer the master equation into the interaction picture. 
Then, we use the relation between the density matrix $\rho$ and the corresponding probability distribution function $\mathcal{P}$ in terms of the Bargmann states \cite{Gardiner1}, given by
\begin{align}
\label{Eqn_A10}
\rho= \int d^{2}\alpha_{j} ||\alpha_{j}\rangle \langle \alpha_{j}||~e^{-\alpha_{j} \alpha^{*}_{j}}\mathcal{P}(\alpha_{j},\alpha^{*}_{j}),
\end{align}
Here $||\alpha_{j}\rangle$ represents a Bargmann state with amplitude $\alpha_{j}$ and employ the following properties representing how the action of the creation and annihilation operators on $\rho$ is reflected  in the space of variables of $\mathcal{P}$:
\begin{align}
b_{j}\rho~\leftrightarrow~\alpha_{j}\mathcal{P}(\alpha_{j},\alpha^{*}_{j}), \nonumber \\
b^{\dagger}_{j}\rho~\leftrightarrow~\left(\alpha^{*}_{j}-\frac{\partial}{\partial\alpha_{j}}\right)\mathcal{P}(\alpha_{j},\alpha^{*}_{j}), \nonumber \\
\rho b_{j}~\leftrightarrow~\left(\alpha_{j}-\frac{\partial}{\partial\alpha^{*}_{j}}\right)\mathcal{P}(\alpha_{j},\alpha^{*}_{j}), \nonumber \\
\rho b^{\dagger}_{j}~\leftrightarrow~\alpha^{*}_{j}\mathcal{P}(\alpha_{j},\alpha^{*}_{j}).
\label{Eqn_A11}
\end{align}
$\mathcal{P}(\alpha_{j},\alpha^{*}_{j})$ is the probability distribution function with $\alpha_{j} = Q_{j}+iP_{j}$ and $\alpha^{*}_{j}$ being its complex conjugate \cite{Gardiner1}. 
Next, for simplicity, we consider the case $A_{t} \gg \gamma_{g},\gamma_{a},s$ and make use of Eqs.~(\ref{Eqn_A10}) and (\ref{Eqn_A11}) to write an approximate FP equation for the coupled system as
\begin{align}
\label{Eqn_A12}
\frac{\partial \mathcal{P}}{\partial t}&\approx \frac{A_{t}}{4}\frac{\partial^{2}}{\partial Q^{2}_{1}}\mathcal{P}+ \gamma_{g}\frac{\partial }{\partial Q_{1}}Q_{1}\mathcal{P} + \frac{A_{t}}{4}\frac{\partial^{2}}{\partial P^{2}_{1}}\mathcal{P} \nonumber \\
                                                 &+ \gamma_{g}\frac{\partial }{\partial P_{1}}P_{1}\mathcal{P}-s\frac{\partial }{\partial Q_{1}}P_{1}\mathcal{P}+s\frac{\partial }{\partial P_{1}}Q_{1}\mathcal{P}\nonumber \\
                                                 &-s\frac{\partial }{\partial P_{1}}Q_{2}\mathcal{P}+s\frac{\partial }{\partial Q_{1}}P_{2}\mathcal{P}\nonumber \\
                                                 &+\frac{A_{t}}{4}\frac{\partial^{2}}{\partial Q^{2}_{2}}\mathcal{P}+ (\gamma_{g}-\gamma_{g}r-\gamma_{a})\frac{\partial }{\partial Q_{2}}Q_{2}\mathcal{P} \nonumber \\
                                                 &+ \frac{A_{t}}{4}\frac{\partial^{2}}{\partial P^{2}_{2}}\mathcal{P}+ (\gamma_{g}+\gamma_{g}r-\gamma_{a})\frac{\partial }{\partial P_{2}}P_{2}\mathcal{P} \nonumber \\
                                                 &+6\gamma_{f}\frac{\partial }{\partial Q_{2}}Q_{2}(Q^{2}_{2}+P^{2}_{2})\mathcal{P}+6\gamma_{f}\frac{\partial }{\partial P_{2}}P_{2}(Q^{2}_{2}+P^{2}_{2})\mathcal{P},
\end{align}
where $\mathcal{P} \equiv \mathcal{P}(Q_{1},P_{1},Q_{2},P_{2})$.
\renewcommand{\theequation}{D-\arabic{equation}}
  \setcounter{equation}{0}  
    \renewcommand{\thefigure}{D-\arabic{figure}}
  \setcounter{equation}{0}
\begin{figure}[t!]
\centering\includegraphics[height=6.5cm, width=8.0cm]{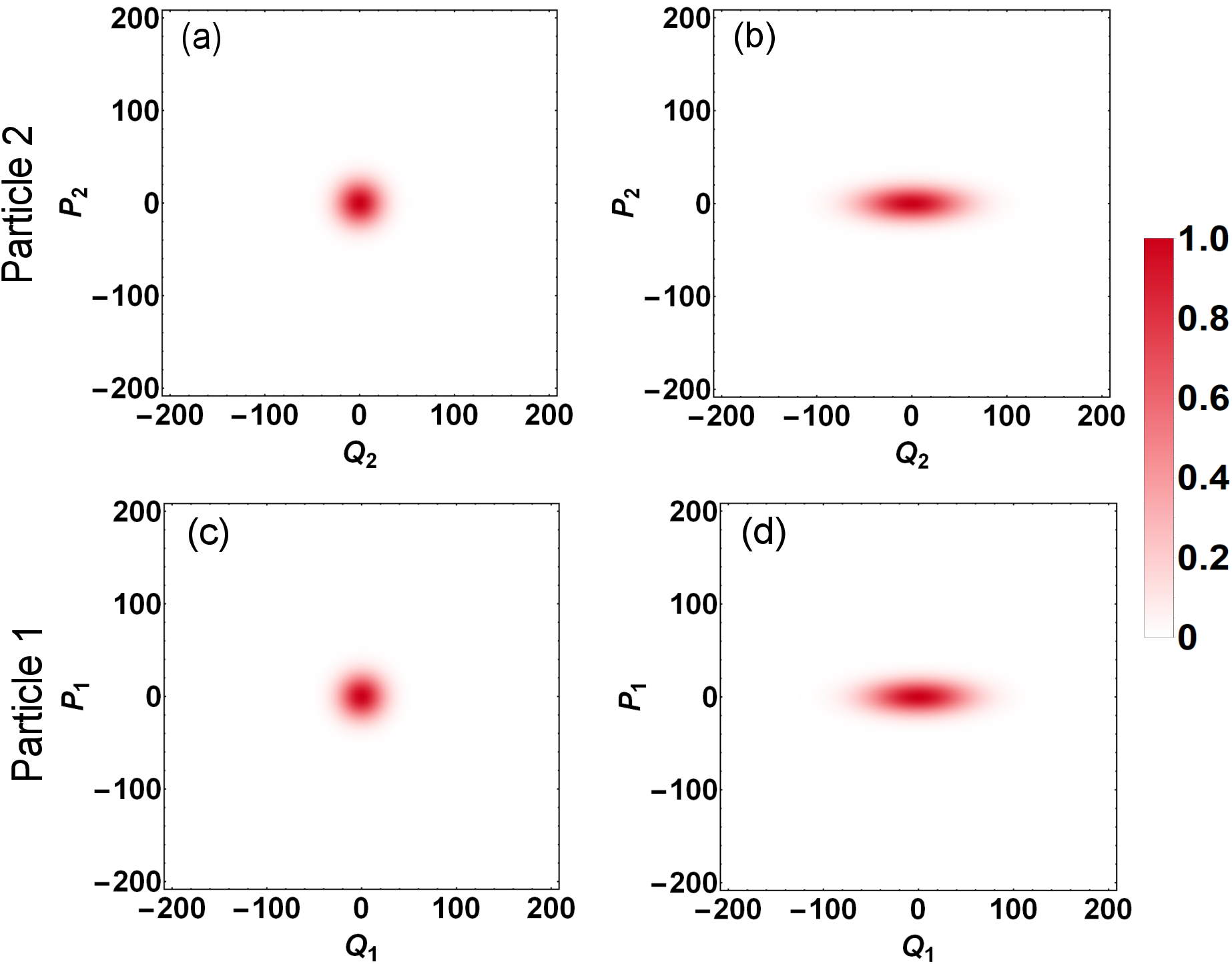}%
 \caption[]{Phase-space distribution of motion of both the nanoparticles plotted using analytical expressions of $\mathcal{P}$ as in Eqs.~(\ref{Eqn_A13}) and (\ref{Eqn_A14}). Panel (a) [(b)] shows the state of nanoparticle 2 before [after] parametric driving. Panel (c) [(d)] represents the state of nanoparticle 1 before [after] coupling. All parameters are the same as in Fig.~\ref{fig:fig2} }
\label{fig:fig6}%
\end{figure}
\section{Analysis for squeezed states}
\label{AppxD}
In the case of the generation of squeezed states, we consider $\gamma_{a}=0$ and $\gamma_{f}=0$ in the above FP equation, Eq.~(\ref{Eqn_A12}), and find an analytical solution for it in the steady-state limit. By assuming $A_{t} \gg \gamma_{g}$, and $s>\gamma_{g}$ and then making use of the drift-diffusion approach for the interacting system as in Ref.\cite{Risken1}, we find such solutions for the probability distribution functions $\mathcal{P}$ of both nanoparticles as
\begin{align}
\label{Eqn_A13}
\mathcal{P}_{2}(Q_{2},P_{2})&\approx C_{2}\mathrm{exp}\left[ -\frac{2}{A_{t}}\Bigl\{(\gamma_{g}-\gamma_{g}r)Q^{2}_{2}+(\gamma_{g}+\gamma_{g}r)P^{2}_{2}\Bigr\}\right] \\
\mathcal{P}_{1}(Q_{1},P_{1})&\approx C_{1}\mathrm{exp}\left[ -\frac{2}{A_{t}}\biggl\{ \Bigl( \gamma_{g}-\gamma_{g}r+ \frac{\gamma^{2}_{g}}{s^{2}}(\gamma_{g}+\gamma_{g}r)\Bigr) Q^{2}_{1}\biggr\}\right] \nonumber \\
                                                 & \times  \mathrm{exp}\left[ -\frac{2}{A_{t}}\biggl\{ \Bigl(\gamma_{g}+\gamma_{g}r+ \frac{\gamma^{2}_{g}}{s^{2}}(\gamma_{g}-\gamma_{g}r)\Bigr)P^{2}_{1}\biggr\}\right] \nonumber \\
                                                 & \times  \mathrm{exp}\left[ -\frac{2}{A_{t}}\frac{4\gamma^{2}_{g}r}{s}Q_{1}P_{1}\right],
\label{Eqn_A14}
\end{align}
where $\mathcal{P}_{2}(Q_{2}, P_{2})$ and $\mathcal{P}_{1}(Q_{1}, P_{1})$ are the reduced probability distribution function for nanoparticle 2 and nanoparticle 1, respectively \cite{Risken1}. Further, $C_{1}$ and $C_{2}$ are the normalization constants.

To quantify the squeezing in the coupled levitated system, we calculate mean values as
\begin{align}
\label{Eqn_A15}
\langle \mathcal{O}\rangle= \frac{\int \mathcal{O} ~\mathcal{P}_{j}(\mathcal{O}) ~d\mathcal{O}}{\int \mathcal{P}_{j}(\mathcal{O})~ d\mathcal{O}},~~~~\mathcal{O} \in \{Q_{2}, P_{2},Q_{1}, P_{1}\},
\end{align}
and derive analytical expressions for the variances of the quadrature components of both the nanoparticles, leading to
\begin{align}
\label{Eqn_A16}
\sigma^{2}_{Q_{2}}&=\langle Q^{2}_{2}\rangle-\langle Q_{2}\rangle^{2} = \frac{A_{t}}{4\gamma_{g}(1-r)}, \\
\sigma^{2}_{P_{2}}&=\langle P^{2}_{2}\rangle-\langle P_{2}\rangle^{2} = \frac{A_{t}}{4\gamma_{g}(1+r)}, \\
\sigma^{2}_{Q_{1}}&=\langle Q^{2}_{1}\rangle-\langle Q_{1}\rangle^{2} \nonumber \\
&= \frac{A_{t}s^{2}}{4\gamma^{2}_{g}(1-r^{2})(\gamma^{2}_{g}+s^{2})^{2}}\left[\gamma^{3}_{g}(1-r)+\gamma_{g}(1+r)s^{2}\right] \nonumber \\
&\approx \frac{A_{t}}{4\gamma_{g}(1-r)}~~~~ \mbox{for} ~~ s \gg \gamma_{g}, \\
\sigma^{2}_{P_{1}}&=\langle P^{2}_{1}\rangle-\langle P_{1}\rangle^{2}  \nonumber \\
&= \frac{A_{t}s^{2}}{4\gamma^{2}_{g}(1-r^{2})(\gamma^{2}_{g}+s^{2})^{2}}\left[\gamma^{3}_{g}(1+r)+\gamma_{g}(1-r)s^{2}\right] \nonumber \\
&\approx \frac{A_{t}}{4\gamma_{g}(1+r)}~~~~ \mbox{for} ~~ s \gg \gamma_{g}.
\label{Eqn_A17}
\end{align}
Finally, to quantify the efficiency of the induced state-transfer process, we evaluate the corresponding fidelity as\cite{Lee1, Sharma2}
\begin{align}
\label{Eqn_A18}
\mathcal{F}= \frac{\iint \mathcal{P}_{1}(Q,P)~\mathcal{P}_{2}(Q,P)~dQ~dP}{\iint [\mathcal{P}_{2}(Q,P)]^{2}~dQ~dP}.
\end{align}
\renewcommand{\theequation}{E-\arabic{equation}}
  \setcounter{equation}{0}  
   \renewcommand{\thefigure}{E-\arabic{figure}}
  \setcounter{equation}{0}  
\section{Analysis for random-phase coherent states}
\label{AppxE}
\begin{figure}[t!]
\centering\includegraphics[height=4.5cm, width=9.0cm]{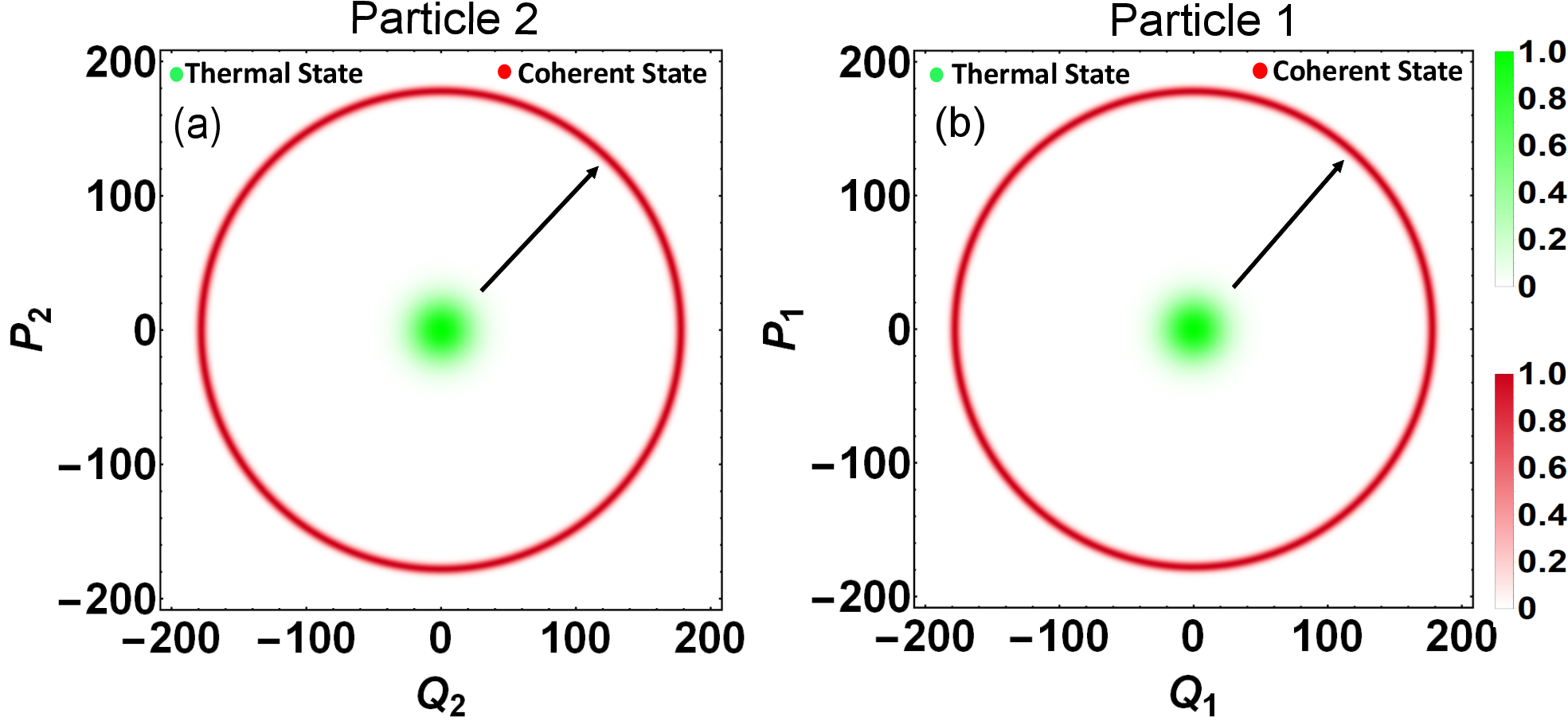}%
 \caption[]{Phase-space distribution of motion of both the nanoparticles plotted using analytical expressions of $\mathcal{P}$ as in Eqs.~(\ref{Eqn_A19}) and (\ref{Eqn_A20}). Panel (a) [(b)] shows the transition from the initial thermal state to a random-phase coherent state of nanoparticle 2 [1]. The arrow depicts the time evolution of the mechanical state. Parameters are the same as in Fig.~\ref{fig:fig3}.}
\label{fig:fig7}%
\end{figure}
For creating and transferring random-phase coherent states, we consider $r=0$ in Eq.~\eqref{Eqn_A12} and use the method as suggested in the above section along with the assumption that $A_{t} \gg \gamma_{g},\gamma_{a}$ and $s>\gamma_{g}$ and derive the analytical solution for the probability distribution function $\mathcal{P}$ of both nanoparticles in the steady-state limit. This gives
\begin{align}
\label{Eqn_A19}
\mathcal{P}_{2}(Q_{2},P_{2})&\approx C_{4}\mathrm{exp}\left[ -\frac{2}{A_{t}}\Bigl\{(\gamma_{g}-\gamma_{a})(Q^{2}_{2}+P^{2}_{2})\Bigr\}\right] \nonumber \\
                                                  &\times \mathrm{exp}\left[ -\frac{2}{A_{t}}\Bigl\{3\gamma_{f}(Q^{2}_{2}+P^{2}_{2})^{2}\Bigr\}\right], \\
\mathcal{P}_{1}(Q_{1},P_{1})&\approx C_{3}\mathrm{exp}\left[ -\frac{2}{A_{t}}\biggl\{(\gamma_{g}-\gamma_{a})\Bigl(1+\frac{\gamma^{2}_{g}}{s^{2}}\Bigr)(Q^{2}_{1}+P^{2}_{1})\biggr\}\right] \nonumber \\
                                                 & \times\mathrm{exp}\left[ -\frac{2}{A_{t}}\biggl\{3\gamma_{f}\Bigl(1+\frac{\gamma^{2}_{g}}{s^{2}}\Bigr)^{2}(Q^{2}_{1}+P^{2}_{1})^{2}\biggr\}\right].
\label{Eqn_A20}
\end{align}
Here, $\mathcal{P}_{2}(Q_{2}, P_{2})$ and $\mathcal{P}_{1}(Q_{1}, P_{1})$ are the reduced probability distribution functions for the nanoparticle 2 and nanoparticle 1, respectively \cite{Risken1} and, $C_{3}$ and $C_{4}$ are the normalization constants. 
By using Eq.~\eqref{Eqn_A15}, we can find the steady-state phonon populations for both nanoparticles as
\begin{align}
\label{Eqn_A21}
N_{2}&=\langle Q^{2}_{2}+P^{2}_{2}\rangle = \frac{\gamma_{a}-\gamma_{g}}{6\gamma_{f}}, \\
N_{1}&=\langle Q^{2}_{1}+P^{2}_{1}\rangle = \frac{(\gamma_{a}-\gamma_{g})s^{2}}{6\gamma_{f}(\gamma^{2}_{g}+s^{2})} \nonumber \\
&\approx  \frac{\gamma_{a}-\gamma_{g}}{6\gamma_{f}}~~~~ \mbox{for} ~~ s \gg \gamma_{g},
\label{Eqn_A22}
\end{align}
where $N_{2}$ ($N_{1}$) represents the steady-state phonon number for nanoparticle 2~(1).

%
%
\nocite{*}


\begin{thebibliography}{99}

  \bibitem{Gieseler1} J. Gieseler, L. Novotny, and R. Quidant, Nature Phys. \textbf{9}, 806–810 (2013).
   
  \bibitem{Romero-Isart1} O. Romero-Isart, New J. Phys. \textbf{19}, 123029 (2017). 
   
  \bibitem{Gonzalez-Ballestero1} C. Gonzalez-Ballestero, M. Aspelmeyer, L. Novotny, R. Quidant, and O. Romero-Isart, Science \textbf{374}, eabg3027 (2021).
   
  \bibitem{Yan1} J. Yan, X. Yu, Z. V. Han, T. Li, and J. Zhang, Photon. Res. \textbf{11}, 600-608 (2023).
   
  \bibitem{Arita1} Y. Arita, E. M. Wright, and K. Dholakia, Optica \textbf{5}, 910-917 (2018).
   
  \bibitem{Rieser1} J. Rieser, M. A. Ciampini, H. Rudolph, N. Kiesel, K. Hornberger, B. A. Stickler, M. Aspelmeyer, and U. Delic, Science \textbf{377}, 987-990 (2022).
   
  \bibitem{Slezak1} B. R. Slezak and B. D’Urso, Appl. Phys. Lett. \textbf{114}, 244102 (2019).
   
  \bibitem{Vijayan1} J. Vijayan, Z. Zhang, J. Piotrowski, D. Windey, F. van der Laan, M. Frimmer, and L. Novotny, Nat. Nanotechnol. \textbf{18}, 49–54 (2023).
   
   \bibitem{Penny1} T. W. Penny, A. Pontin, and P. F. Barker, Phys. Rev. Research \textbf{5}, 013070 (2023). 
   
   \bibitem{Bykov1} D. S. Bykov, L. Dania, F. Goschin, and T. E. Northup, Optica \textbf{10}, 438-442 (2023).   
   
   \bibitem{Liska1} V. Liška, T. Zemánková, V. Svak, P. Jákl, J. Ježek, M. Bránecký, S. H. Simpson, P. Zemánek, and O. Brzobohatý, Optica \textbf{10}, 1203-1209 (2023).
   
  \bibitem{Vijayan2} J. Vijayan, J. Piotrowski, C. Gonzalez-Ballestero, K. Weber, O. Romero-Isart, and Lukas Novotny, Nat. Phys. (2024). {\it DOI: https://doi.org/10.1038/s41567-024-02405-3}.
   
  \bibitem{Rudolph1} H. Rudolph, K. Hornberger, and B. A. Stickler, Phys. Rev. A \textbf{101}, 011804(R) (2020).
   
  \bibitem{Chauhan1} A. K. Chauhan, O. Černotík, and R. Filip, New J. Phys. \textbf{22}, 123021 (2020). 
   
   \bibitem{Brandao1} I. Brandao, D. Tandeitnik, and T. Guerreiro, Quantum Sci. Technol. \textbf{6}, 045013 (2021).  
   
   \bibitem{Rudolph2} H. Rudolph, U. Delić, M. Aspelmeyer, K. Hornberger, and B. A. Stickler, Phys. Rev. Lett. \textbf{129}, 193602 (2022).
   
   \bibitem{Yokomizo1} K. Yokomizo and Y. Ashida, Phys. Rev. Research \textbf{5}, 033217 (2023). 
   
  \bibitem{Reisenbauer1} M. Reisenbauer, H. Rudolph, L. Egyed, K. Hornberger, A. V. Zasedatelev, M. Abuzarli, B. A. Stickler, and U. Delić, (2023), arXiv:2310.02610.
   	
   \bibitem{Liska2} V. Liška, T. Zemánková, P. Jákl, M. Šiler, S. H. Simpson, P. Zemánek, and O. Brzobohatý, (2023), arXiv:2310.03701.   

   \bibitem{Groblacher1} S. Gröblacher, K. Hammerer, M. R. Vanner, and  M. Aspelmeyer, Nature \textbf{460}, 724–727 (2009).   
   	
   \bibitem{Verhagen1} E. Verhagen, S. Deléglise, S. Weis, A. Schliesser, and  T. J. Kippenberg, Nature \textbf{482}, 63–67 (2012).   
   	
  \bibitem{Xu1} H. Xu, L. Jiang, A. A. Clerk, and J. G. E. Harris, Nature \textbf{568}, 65-69 (2019).   	   
   
   \bibitem{Sharma1} S. Sharma, A. Kani, and M. Bhattacharya, Phys. Rev. A \textbf{105}, 043505 (2022).
   
   \bibitem{Rudolph3} H. Rudolph, U. Delić, K. Hornberger, and B. A. Stickler, (2023), arXiv:2306.11893.
   	
   \bibitem{Pettit1} R. M. Pettit, W. Ge, P. Kumar, D. R. Luntz-Martin, J. T. Schultz, L. P. Neukirch, M. Bhattacharya, and A. N. Vamivakas, Nat. Photonics \textbf{13}, 402–405 (2019).
    	
    \bibitem{Bothner1} D. Bothner, S. Yanai, A. Iniguez-Rabago, M. Yuan, Ya.M. Blanter, and G.A. Steele, Nat. Commun. \textbf{11}, 1589 (2020).
   	
   \bibitem{Mahboob1} I. Mahboob, H. Okamoto, K. Onomitsu, and H. Yamaguchi, Phys. Rev. Lett. \textbf{113}, 167203 (2014).
   
    \bibitem{Pontin1} A. Pontin, M. Bonaldi, A. Borrielli, F. S. Cataliotti, F. Marino, G. A. Prodi, E. Serra, and F. Marin, Phys. Rev. Lett. \textbf{112}, 023601 (2014).
   
    \bibitem{Aspelmeyer1} M. Aspelmeyer, T. J. Kippenberg, and F. Marquardt, Rev. Mod. Phys. \textbf{86}, 1391-1452 (2014).
   
    \bibitem{Neto1} G. D. de Moraes Neto, F. M. Andrade, V. Montenegro, and S. Bose, Phys. Rev. A \textbf{93}, 062339 (2016).   
   
     \bibitem{Xi1} H. Xi and P. Pei, Phys. Rev. A \textbf{104}, 052421 (2021).
      
   \bibitem{Navarathna1} A. Navarathna, J. S. Bennett, and W. P. Bowen, Phys. Rev. Lett. \textbf{130}, 263603 (2023).   
   
    \bibitem{Matsuo1} K. Matsuo, Phys. Rev. A \textbf{41}, 519 (1990). 
   
    \bibitem{Riedinger1} R. Riedinger, A. Wallucks, I. Marinkovic, C. Loschnauer, M. Aspelmeyer, S. Hong, and S. Groblacher, Nature \textbf{556}, 473–477 (2018).   
   
    \bibitem{Degen1} C. L. Degen, F. Reinhard, and P. Cappellaro, Rev. Mod. Phys. \textbf{89}, 035002 (2017).    
   
   \bibitem{Kwon1} H. Kwon, H. Jeong, D. Jennings, B. Yadin, and M. S. Kim, Phys. Rev. Lett. \textbf{120}, 150602 (2018).    
   
   \bibitem{Manzano1} G. Manzano, J. M. R. Parrondo, and G. T. Landi, PRX Quantum \textbf{3}, 010304 (2022).  
   
   \bibitem{Chu1} Y. Chu, X. Li, and J. Cai, Phys. Rev. Lett. \textbf{130}, 170801 (2023). 
   
   \bibitem{Ricci2} F. Ricci, R. A. Rica, M. Spasenovic, J. Geiseler, L. Rondin, L. Novotny, and R. Quidant, Nat. Commun. \textbf{8}, 15141 (2017).
   	
  \bibitem{Landa1} H. Landa, M. Schiro, and G. Misguich, Phys. Rev. Lett. \textbf{124}, 043601 (2020). 
   
   \bibitem{Foss-Feig1} M. Foss-Feig, P. Niroula, J. T. Young, M. Hafezi, A. V. Gorshkov, R. M. Wilson, and M. F. Maghrebi, Phys. Rev. A \textbf{95}, 043826 (2017). 
   
    \bibitem{Gardiner1}C.W. Gardiner, P. Zoller, {\it Quantum Noise: A Handbook of Markovian and Non-Markovian Quantum Stochastic Methods with Applications to Quantum Optics,} (Springer, 2000).
   
   \bibitem{Risken1} H. Risken, {\it The Fokker-Planck Equation: Methods of Solution and Applications,} (Springer-Verlag Berlin Heidelberg, 1989).
  
    \bibitem{Lee1}J. Lee, M. S. Kim, and H. Jeong, Phys. Rev. A \textbf{62}, 032305 (2000). 
  
   \bibitem{Sharma2} S. Sharma and M. Bhattacharya, J. Opt. Soc. Am. B \textbf{37}, 1620-1629 (2020).



\end{thebibliography}

\end{document}